\documentclass [12pt,a4paper]{article}
\usepackage{feynmf}
\newcommand{\be}{\begin{equation}} \newcommand{\ee}{\end{equation}}
\newcommand{\ba}{\begin{eqnarray}} \newcommand{\ea}{\end{eqnarray}}
\newcommand{\bes}{\begin{equation*}} \newcommand{\ees}{\end{equation*}}
\newcommand{\bas}{\begin{eqnarray*}} \newcommand{\eas}{\end{eqnarray*}}

\begin{document}

\centerline{\bf\Large Cosmological Symmetry Breaking, Pseudo-scale }
\centerline{\bf\Large invariance, Dark Energy and the Standard Model}

\bigskip
\centerline{\bf \large Pankaj Jain and Subhadip Mitra}

\bigskip
\centerline{Physics Department, I.I.T. Kanpur, India 208016}

\bigskip
\noindent
{\bf Abstract:} The energy density of the universe today may be dominated by
the vacuum energy of a slowly rolling scalar field. 
Making a quantum expansion around such a time dependent solution
breaks fundamental symmetries of quantum field theory. 
We call this mechanism
cosmological symmetry breaking and argue that it is different from the
standard phenomenon of spontaneous symmetry breaking. 
We illustrate this with a toy scalar field
theory, whose action displays a U(1) symmetry. We identify a
symmetry, called pseudo-scale invariance, which sets the cosmological
constant exactly equal to zero,
both in classical and quantum theory. This symmetry is also broken
cosmologically and leads to a nonzero vacuum or dark energy. The slow
roll condition along with the observed value of dark energy leads to a value
of the background scalar field of the order of Planck mass. We also consider
a U(1) gauge symmetry model. Cosmological symmetry breaking, in this case,
leads to a non zero mass for the vector field. We also show that a
cosmologically broken pseudo-scale
invariance can generate a wide range of masses.

\section{Introduction}
The current cosmological observations
\cite{Riess1998,Garnavich,Perlmutter,Tonry,Barris,Riess2004}
suggest that the energy
density of the universe
gets a significant contribution from vacuum energy. For a review see
\cite{Peebles,Padmanabhan,Copeland}. This may be modelled
by simply introducing a cosmological constant
\cite{Peebles,Padmanabhan,Weinberg,Carroll,Sahni,Ellis} or dynamically
by a scalar field slowly
rolling towards the true minimum of the potential
\cite{Caldwell}. If we assume the existence of such a field
then it implies that in the current era its lowest energy state is not the
true vacuum state of the theory. In order to study the spectrum of this theory one needs
to make a quantum expansion around a time dependent field. This has interesting
implications for the physics of such models, not explored so far in the
literature. In particular since we are expanding around a time dependent
field and not the ground state, the resulting physics need not display the
symmetries of the original lagrangian. Hence this may provide us with another
method of breaking fundamental symmetries. We point out that in general
a theory displays the symmetries of the action provided the ground state
of the theory is symmetric under the corresponding transformations.
In the present case, however, the ground state is irrelevant since the field
never reaches this state. The time dependent
solution, around which we are required to make the quantum expansion, need not
display the symmetries of the original action. Hence the theory may display
broken symmetries over a large period in the life time of the universe.

\section{ Symmetry Breaking for a Scalar Field}
We consider a simple model of a complex scalar field with a global U(1)
symmetry
\begin{eqnarray}
\mathcal{L}[\Phi(x)]=\partial_{\mu}\Phi^{*}(x)\partial^{\mu}\Phi(x)
-m^2\Phi^{*}(x)\Phi(x)-\lambda(\Phi^{*}(x)\Phi(x))^2.
\end{eqnarray}
This Lagrangian has global $U(1)$ symmetry, i.e., it is invariant
under $\Phi(x)\rightarrow e^{i\theta}\Phi(x)$.
So far we have neglected the effect of the background metric which will also
be included later.
We assume that the potential
is sufficiently gentle, with mass parameter sufficiently small,
such that the scalar field is slowly rolling towards its true minimum.
Let $\eta(t)$ be the classical solution to the equations of motion. Here
we assume that this solution is independent of space and is in general complex.

We split $\Phi(x)$
into two parts: $\Phi(x)= \eta(x)+ \phi(x)$. Here $\eta(x)$ is the
classical solution of the equation of motion. Let us assume the
classical solution depends only on time but is in general complex, i.e,
$\eta(x) = \eta(t)$. Hence $\Phi(x)=\eta(t)+ \phi(x)$ and
$\Phi^{*}(x)=\eta^*(t)+\phi^{*}(x)$.
\begin{eqnarray}
\mathcal{L}[\eta,\phi] &=&  \dot{\eta^*}\dot{\eta} +
\dot{\eta^*}\dot{\phi}+\dot{\eta}\dot{\phi^{*}} +
\partial_{\mu}\phi^{*}\partial^{\mu}\phi - m^2(\phi^{*}\phi + \eta\phi^{*} +
\eta^*\phi + \eta^*\eta)\nonumber\\
 &-&\lambda\{(\phi^{*}\phi)^2 + (\eta\phi^{*})^2 + (\phi\eta^*)^2
+ (\eta^*\eta)^{2} + 2(\eta\phi^{*} +
\eta^*\phi)\phi^*\phi\nonumber\\
&+&4\eta^*\eta\phi^{*}\phi
+ 2\eta^*\eta(\eta^*\phi+\eta\phi^*)\}.
\end{eqnarray}
From this we identify the classical Lagrangian.
\begin{eqnarray}
\mathcal{L}_{Classical}[\eta(t)] =  \dot{\eta^*}\dot\eta-
m^2\eta^*\eta-\lambda(\eta^*\eta)^2.
\end{eqnarray}
The classical field $\eta$ satisfies the equation of motion,
\begin{eqnarray}
\ddot{\eta}+\eta(m^2+2\lambda\eta^*\eta)=0.
\end{eqnarray}
If we assume that the quantum fluctuations die sufficiently fast at t =
$\pm\infty$, then we can write
\begin{eqnarray}
\nonumber \int d^{4}x\;\dot{\eta}(\dot{\phi}+\dot{\phi^{*}})=-\int
d^{4}x\;\ddot{\eta}(\phi+\phi^{*})= \int
d^{4}x\;\eta(m^2+2\lambda\eta^2)(\phi+\phi^{*}).
\end{eqnarray}
Hence we get,
\begin{eqnarray}
\nonumber \mathcal{L} &=& \mathcal{L}_{Classical}[\eta(t)]
+\partial_{\mu}\phi^{*}\partial^{\mu}\phi - m^2\phi^{*}\phi\\ &-&
\lambda\{(\phi^{*}\phi)^2
+ (\eta\phi^{*})^2 + (\eta^*\phi)^2 + 2\phi^{*}\phi(\eta\phi^{*} +
\eta^*\phi)
+4\eta^*\eta\phi^{*}\phi\}.
\end{eqnarray}
Let $\eta = \eta_0 e^{i\theta_0}$ and $\phi(x)=(\phi_1 + i\phi_2)/\sqrt{2}$.
This gives
\begin{eqnarray}
\nonumber \mathcal{L} &=&\mathcal{L}_{Classical}[\eta(t)]+
{1\over 2}\partial_{\mu}\phi_1\partial^{\mu}\phi_1 + {1\over 2}
\partial_\mu\phi_2\partial^\mu\phi_2
- {m^2\over 2}(\phi_1^2+\phi_2^2)\nonumber\\
&-& {\lambda\eta_0^2\over 4}[2(\phi_1^2-\phi_2^2)\cos2\theta_0
+ 4\phi_1\phi_2 \sin2\theta_0 + 4(\phi_1^2 + \phi_2^2)]\nonumber\\
&-&{\lambda\over 4}[(\phi_1^2 + \phi_2^2)^2+4\eta_0(\phi_1^2 + \phi_2^2)(\phi_1
\cos\theta_0 + \phi_2\sin\theta_0)].
\end{eqnarray}
It is clear that the two modes do not have the same mass. For a complex
classical solution, i.e. with $\theta_0\ne 0$, the two modes are coupled.
They can be decoupled by the rotation in internal space
\begin{equation}
\left(\matrix{\phi_1\cr \phi_2}\right) = \left(\matrix{\cos\beta &-\sin\beta\cr
\sin\beta & \cos\beta}\right)\left(\matrix{\phi'_1\cr \phi'_2}\right).
\end{equation}
We assume adiabaticity and hence we ignore the time dependence of
rotation matrix in the kinetic energy term. The rotation angle $\beta$ is
found to be equal to $\theta_0$ and the two mass eigenvalues are found
to be $m^2+6\lambda\eta_0^2$ and $m^2+2\lambda\eta_0^2$. Hence the two modes
pick up different masses and the symmetry is broken. We do not find
a zero mode, in contrast to the case of spontaneous symmetry breaking.
We call this phenomenon Cosmological Symmetry Breaking.

\section{Gravitational Background}
We next consider the complex scalar field model in a background gravitational
field, which would be considered classically. The action may be written as
\begin{eqnarray}
\mathcal{S}=\int d^4x\sqrt{-g}\left[g^{\alpha\beta}
\partial_{\alpha}\Phi^{*}(x)\partial_{\beta} \Phi(x) -
m^2\Phi^{*}(x)\Phi(x)-\lambda(\Phi^{*}(x)\Phi(x))^2\right].
\end{eqnarray}
We assume the FRW background
metric with the Hubble parameter $H(t)$ and the expansion factor $R(t)$. This
model also displays the global U(1) symmetry $\Phi(x)\rightarrow
e^{i\theta} \Phi(x)$. We again write $\Phi(x)= \eta(t)+ \phi(x)$, where
$\eta(t)$ is a space independent solution to the classical equations of motion.
The real and
imaginary parts  $\eta_1$ and $\eta_2$ satisfy
\begin{equation}
{d^2\eta_i\over dt^2} + 3 H {d\eta_i\over dt} + {\partial V\over\partial
\eta_i}=0
\end{equation}
for $i=1,2$.
Here the potential $V(\eta) = m^2\eta^*\eta + \lambda (\eta^*\eta)^2$.
The model again displays symmetry breaking as long as $\eta(t)$ is different
from zero. This is allowed as long as the conditions for slow roll is satisfied.
The Hubble parameter is determined by the entire matter content of
the universe and here we shall consider it as an independent function of
$t$. We shall assume it to be approximately constant as is the case for a
vacuum dominated universe.
The slow roll condition is satisfied if the mass parameter $m<<3H$.
We may
expand the potential around the classical solution and find that the
mass spectrum is same as found in the earlier section. We, therefore, find
a squared mass splitting $4\lambda \eta(t)^2$, whose scale is determined by
the Hubble parameter. Since the splitting is determined by the value
of the Hubble parameter, its value in the current era is quite small.
However the splitting need not be small in the early universe.
At that time it may lead to large observable consequences.

Although the mass splitting is very small, the value of the field $\eta(t)$
can be large. This depends on our choice of the coupling $\lambda$. By
choosing $\lambda$ sufficiently small we can make $\eta(t)$ arbitrarily
large and still maintain slow roll conditions. The non-zero value of this
field can lead to a wide range of breakdown of symmetries, including Lorentz
invariance.  Lorentz invariance is broken because the classical field
only has time
dependence. Furthermore
if we gauge the U(1) invariance then the gauge symmetry will be
broken. The mass of the gauge field in this case depends on $\eta(t)$ and
can be quite large.

\section{Dark Energy}

So far we have considered a slowly rolling complex scalar field and shown
that it leads to breakdown of symmetries of the original lagrangian.
We next consider the possibility that the complex scalar field
itself leads to dark energy. We determine the range of allowed
values for the background scalar field in order that it leads to vacuum
energy equal to the observed dark energy density. For this purpose we
consider the equation of motion in gravitational background in terms of the
two real fields $\eta_1$ and $\eta_2$. For orders of magnitude estimate we may
assume $\eta_1\sim \eta_2$. The analysis is easily modified if this is not
the case and does not lead to any essential difference. For slow roll,
the second derivative term is negligible. We may consider two separate
cases where either mass term or the quartic coupling term in the potential
dominates. If the mass term dominates then the slow roll condition is
satisfied if $m^2<< 9 H^2$. If the quartic coupling term dominates then
we find the condition $\lambda << H^2/\eta^2$. We next require that
$\rho_V = V(\eta)$. In both the cases this leads to the condition,
\begin{equation} \eta >> \rho_V^{1/2}/H \approx \sqrt{3\over 8\pi}\, M_{\rm PL}
\end{equation}
where $M_{\rm PL}$ is the Planck mass.
Here we have used the fact that the vacuum energy density is almost equal
to the
critical energy density. Hence we find that
in our model the value of
the slow roll scalar field has to be of the order of the Planck mass or higher.

We point out that the value of the coupling constant $\lambda$ turns out
to be extremely small in our model for dark energy. This by itself need not
lead to a fine tuning problem since we are free to choose a value for
this parameter. Indeed a parameter as small as this is expected due to the
widely different scales of Planck mass and the Hubble constant. The fine
tuning problem may arise if at higher orders we need to adjust this parameter
to very high accuracy. This may happen if it undergoes large quantum
corrections. This is an important check of the theory which we shall
address in detail
in a future publication.

\section{Gauge Theory}
We next gauge the U(1) symmetry considered in the earlier sections. The
resulting Lagrangian, in the presence of background gravity, can be written as
\ba
\nonumber \mathcal{S}&=&\int d^4x\sqrt{-g}\left[g^{\alpha\beta}
(\mathcal{D}_{\alpha}\Phi(x))^{*}\mathcal{D}_{\beta} \Phi(x) -\frac14 g^{\alpha\beta}g^{\kappa\delta}
\mathcal{F}_{\alpha\kappa}\mathcal{F}_{\beta\delta} -
m^2\Phi^{*}(x)\Phi(x)\right.\\
&-&\left.\lambda(\Phi^{*}(x)\Phi(x))^2\right]
\ea
where $\mathcal{D}_\alpha = \partial_\alpha -ig\mathcal{A}_\alpha$ is the covariant derivative
and $\mathcal{F}_{\alpha\beta}$ the field strength tensor of the U(1) gauge field
$\mathcal{A}_\alpha$.
This Lagrangian is invariant under $\Phi(x)\rightarrow
e^{i\theta(x)}\Phi(x)$. We can parameterize $\Phi(x)$ by:
\bas
\nonumber \Phi(x) &=& (\eta_0 + \rho/\sqrt{2})\exp{\left[
i\left(\theta_0+{\sigma\over \sqrt{2}\eta_0}\right)\right]}\\
&=& \left(\eta_0 + {\rho\over \sqrt{2}} + i{\sigma\over \sqrt{2}} +\dots\right)
e^{i\theta_0}.
\eas
Hence, for small oscillations $\rho(x)$ and $\sigma(x)$ are $\phi'_1(x)$
and $\phi'_2(x)$ respectively. We define new fields,
\bas
\Phi' &=& \exp{\left[-i\left(\theta_0+{\sigma\over\sqrt{2}\eta_0}\right)\right]}\Phi = \eta_0 +\rho/\sqrt{2},\\
\mathcal{B}_\mu &=& \mathcal{A}_\mu - \frac1g\partial_\mu\left(\theta_0+\frac{\sigma}{\sqrt{2}\eta_0}\right).
\eas
If we neglect the gravitational field, the Lagrangian becomes:
\bas
\mathcal{L} =(\mathcal{D}_{\mu}\Phi(x))^{*}(\mathcal{D}^{\mu}\Phi(x))
-\frac14\mathcal{F}_{\mu\nu}\mathcal{F}^{\mu\nu} -
m^2(\eta_0+\rho/\sqrt{2})^2-\lambda(\eta_0+\rho/\sqrt{2})^4.
\eas

We now follow the same procedure as before. We split the gauge field into two parts
 - $i)$ the classical gauge field, $\beta_{\mu}(t)$
which depends only on time and $ii)$ $B_{\mu}(x)$, the quantum
field i.e., $\mathcal{B}_{\mu}(x)=\beta_{\mu}(t)+ B_{\mu}(x)$.
So the classical Lagrangian becomes:
\ba
\nonumber \mathcal{L}_{Classical}&=&\dot{\eta}_0^2
+g^2\beta_{\mu}\beta^{\mu}\eta_0^2-m^2\eta_0^2
-\lambda\eta_0^4-\frac14f_{\mu\nu}f^{\mu\nu}
\ea
where $f_{\mu\nu}=\partial_\mu\beta_\nu-\partial_\nu\beta_\mu$. The equations of motion are:
\ba
\beta_0 &=& 0,\\
\ddot{\beta}_i &=& - 2g^2\beta_i\eta_0^2, \\
\ddot{\eta_0} &=& - \eta_0(g^2\vec{\beta}^2 + m^2 +2\lambda\eta_0^2).
\ea
If we assume $\beta_\mu=0$, then using these equations and dropping the total derivative terms
one can rewrite the Lagrangian as:
\ba
\nonumber \mathcal{L} &=&   \mathcal{L}_{Classical} +
{1\over 2}\partial_\mu\rho\partial^\mu\rho
+ g^2B^2\left(\eta_0^2 +{\rho^2\over 2} +\sqrt{2}\rho\eta_0\right)\\
&-& (m^2+6\lambda\eta_0^2){\rho^2\over 2}
-{\lambda\over 4}(\rho^4 + 4\sqrt{2}\rho^3\eta_0)-\frac14F_{\mu\nu}F^{\mu\nu}
\ea
where $F_{\mu\nu}=\partial_\mu B_\nu - \partial_\nu B_\mu$. The gauge field
has acquired a mass, $m_B = g\eta_0$. Hence the gauge invariance is broken
in this theory. In the simplest case, discussed in the last section, this mass
will be of the order of Planck mass, assuming gauge coupling of order unity
and if we require that the scalar field vacuum energy gives dominant
contribution to dark energy. However if we do not impose the condition that
the vacuum energy associated with the field $\Phi$ is equal to the observed
vacuum energy, then the mass of the gauge field is an independent parameter
which can be fixed by a suitable choice of the classical solution.
In section 7 below we also provide another generalization of the lagrangian
so as to generate a different mass scale.

\section{Pseudo-scale Invariance}
We have so far introduced a new method of breaking symmetries of a field
theory. The procedure is found to naturally lead to dark energy in the form
of the vacuum energy of a slowly rolling scalar field. However so far we
not addressed the question of why the cosmological constant is so small.
The problem is ofcourse well known. Quantum field theory in general produces
cosmological constant many orders of magnitude larger than what is observed.
In the absence of any symmetry, which may demand absence of cosmological term
in the action, this is a serious problem in fundamental physics. In this section
we identify a symmetry which eliminates cosmological constant both at
classical and quantum level.

We first consider actions which are scale invariant. It is clear that at the
classical level scale invariance eliminates all dimensionful parameters from the
action, including a cosmological constant term. However in the quantum theory,
it is well known that scale invariance is anomalous and hence may generate
a cosmological constant. In very interesting papers Cheng \cite{ChengPRL}
and Cheng and Kao \cite{Cheng,Kao} have argued
that scale transformations can be broken into a general coordinate transformation and what is refered to as the pseudo-scale transformations. Under the
pseudo-scale transformations
\ba
x & \rightarrow & x \cr
\Phi &\rightarrow & \Phi/\Lambda\cr
g^{\mu\nu} &\rightarrow & g^{\mu\nu}/\Lambda^2 \cr
A_\mu &\rightarrow & A_\mu.
\ea
The matter part of the action is invariant under this transformation.
The gravitational action is not invariant but, as explained in
Ref. \cite{ChengPRL,Cheng}, it
can be easily generalized so that it is invariant. One simply replaces
\cite{Cheng},
\begin{equation}
{1\over 4\pi G} R \rightarrow \beta \Phi^*\Phi R
\end{equation}
where $G$ is the gravitational constant, $R$ the Ricci scalar and
$\beta$ a dimensionless constant. 
Next assuming a slow role scalar field discussed in earlier sections, with
its value of order the Planck mass, we find that the resulting action
will have predictions identical to Einstein's gravity, at leading order.
The cosmological constant term is not invariant under pseudo-scale invariance
and hence eliminated in the classical action. The theory
now has exactly zero cosmological constant as long as the symmetries of
the action are not broken.

The pseudo-scale invariance is, however, broken through our cosmological
symmetry breaking mechanism, discussed earlier. Hence this mechanism will
generate a nonzero cosmological constant, or dark energy. We can directly
borrow the results obtained in section 4 with the mass parameter $m$ set to
zero. The theory discussed in section 4 now displays pseudo-scale invariance,
besides invariance under the U(1) transformations. Both of these symmetries
are broken cosmologically and the slow roll condition, along with the value
of the observed dark energy density sets the scale of the classical
scalar field of the order of Planck mass.

We next
consider a regulated action, assuming dimensional regularization. Here we 
focus only on scalar fields. In this case we consider the 
scalar field action
in $n$ dimensions,
\begin{eqnarray}
\mathcal{S}=\int d^nx\sqrt{-g}\left[g^{\alpha\beta}
\partial_{\alpha}\Phi^{*}(x)\partial_{\beta} \Phi(x) -
\lambda (\sqrt{-g})^{(n-4)/n} (\Phi^{*}(x)\Phi(x))^2\right].
\end{eqnarray}
The action is invariant under the transformation
\ba
x & \rightarrow & x \cr
\Phi &\rightarrow & \Phi/\Lambda^{a(n)}\cr
g^{\mu\nu} &\rightarrow & g^{\mu\nu}/\Lambda^{b(n)}
\ea
where $b(n) = 4a(n)/(n-2)$ and we may choose $a(n)$ to be any function of
$n$. This generalizes
the pseudo-scale transformations to $n$ dimensions. The action displays
exact symmetry under pseudo-scale transformations in $n$ dimensions. However
the action has general coordinate invariance only in 4 dimensions. In
dimensions other than 4 the potential term violates general coordinate
invariance. Here
we take the point of view that the fundamental quantum theory may obey a more
general transformation law rather than general coordinate invariance. We
are guided primarily by data and the absence of general coordinate invariance
in dimensions other than four
will give modified predictions only at very high energy scale of the order
of Planck mass. At this scale the theory is so far untested and we cannot
rule out our action.

To summarize, we find that we can impose pseudo-scale invariance as
an exact symmetry. The symmetry is not anomalous. This symmetry prohibits
us to introduce a cosmological constant, both at the classical and quantum
level. Hence it may provide an explanation for why the cosmological
constant is so small. Alternative approaches to solve the
cosmological constant problem are described in Ref. 
\cite{Weinberg,Aurilia,VanDer,Henneaux,Brown,Buchmuller,Henneaux89,Sorkin,Sundrum}.

\section{Generating Masses}
The basic problem of generating realistic masses of the observed particles,
 however, still remains in our theory. Pseudo-scale invariance prohibits any
mass terms in the action. Hence the standard Higgs mechanism is not applicable
and all the Standard Model fields, for example, will remain massless.
The pseudo-scale invariance is ofcourse broken cosmologically and hence one
may expect that we may be able to break the standard model gauge symmetry
also by a slowly rolling scalar field. However we have to do this such
that the mass of the scalar field is sufficiently large and not ruled out
experimentally. In the construction so far, the mass of the scalar field
has been found to be very small. One possibility is that this scalar boson
is eliminated from the spectrum by gauging the pseudoscale invariance
\cite{ChengPRL,Cheng}. In this case the Higgs boson will be eliminated
from the spectrum. An alternate construction, which does not involve gauging
the pseudo-scale invariance, is described below.

We next construct a toy model such that, besides generating dark energy, it
also breaks another $U(1)$ symmetry with a sufficiently large mass
of the scalar field. This is a toy model which can be generalized to
construct an acceptable Standard Model of particle physics.
We consider a model with two complex scalar fields $\Phi$ and $\Psi$.
Here $\Phi$ will be considered as a slowly rolling field which gives rise
to dark energy. We construct an action such that it is invariant under the
transformation $\Phi\rightarrow e^{i\theta}\Phi$ and
$\Psi\rightarrow e^{i\xi}\Psi$ as well as the pseudo-scale transformations.
Here we restrict ourselves to four dimensions.
\begin{eqnarray}
\mathcal{S}&=&\int d^4x\sqrt{-g}\Bigg[g^{\alpha\beta}
\partial_{\alpha}\Phi^{*}(x)\partial_{\beta} \Phi(x) +
g^{\alpha\beta}
\partial_{\alpha}\Psi^{*}(x)\partial_{\beta} \Psi(x)
-\lambda(\Phi^{*}(x)\Phi(x))^2 \cr
&-& \lambda_1(\Psi^*\Psi - \lambda_2^2\Phi^*\Phi)^2
\Bigg].
\end{eqnarray}
Here $\lambda_1$ is taken to be of order unity. Its precise value will be
fixed by the mass of the $\Psi$ particle. The coupling $\lambda_2<<1$ and
will be fixed by the magnitude of the classical solution of the field $\Psi$,
which is eventually determined by the scale of symmetry breaking of the
U(1) group of transformation $\Psi\rightarrow e^{i\xi}\Psi$.

We again expand these two fields as $\Phi=\eta(t) + \phi$ and
$\Psi=\zeta(t) + \psi$, where $\eta(t)$ and $\zeta(t)$ are the time dependent
classical fields and $\phi$ and $\psi$ are the quantum fluctuations. The
classical fields satisfy,
\begin{equation}
{d^2\eta_i\over dt^2} + 3H{d\eta_i\over dt} + \lambda (\eta_1^2 + \eta_2^2)
\eta_i + \lambda_1\lambda_2^2[\zeta_1^2+\zeta_2^2 - \lambda_2^2
(\eta_1^2 + \eta_2^2)]\eta_i = 0,
\label{eq:etai}
\end{equation}

\begin{equation}
{d^2\zeta_i\over dt^2} + 3H{d\zeta_i\over dt} +
\lambda_1[\zeta_1^2+\zeta_2^2 - \lambda_2^2
(\eta_1^2 + \eta_2^2)]\zeta_i = 0.
\label{eq:zetai}
\end{equation}
We consider a slow roll solution such that all the second derivative terms
are negligible. We set
$\zeta_i = (1+\delta)\lambda_2\eta_i$,
where $\delta<<1$. We can determine $\delta$ perturbatively by solving the
differential equations. As we have seen earlier, slow roll condition requires
that $\lambda<< H^2/\eta^2$. Here we have an additional term proportional to
$\lambda_1$ in the equation of motion for $\eta_i$, eq. \ref{eq:etai}.
Substituting $\zeta_i
= (1+\delta)\lambda_2\eta_i$ in eq. \ref{eq:zetai}, we get an estimate of
$d\eta_i/dt$. Substituting this in eq. \ref{eq:etai} we find
\begin{equation}
\delta\approx {\lambda \over 2\lambda_1\lambda^2_2}\,.
\end{equation}
We want to choose $\lambda_2$ such that $\zeta_i$ is of order of the Weinberg
Salam symmetry breaking scale. It is clear that in this case $\delta<<1$,
which is required for the self consistency of the perturbative solution.

We can now determine the contribution of the term proportional to $\lambda_1$
to the equation for $\eta_1$. We find that it gives a contribution of order
$\lambda_2^2\lambda(\eta_1^2+\eta_2^2)\eta_i$, which is much smaller
compared to leading order term $\lambda(\eta_1 ^2+\eta_2^2)\eta_i$. Hence
the term proportional to $\lambda_1$ can be treated perturbatively. We also
find the $\lambda_1$ term gives a correction of order
$(\lambda/\lambda_1) \lambda\phi^4$ to the vacuum energy. Since $\lambda/\lambda_1<< 1$, this correction is negligible.
We, therefore, find that the
new term in the lagrangian, proportional to $\lambda_1$, can be ignored
at the leading order in the equation of motion for $\eta_1$ and also gives
negligible correction to the vacuum energy density. Furthermore the complete
solution in its presence can be determined by treating this term
perturbatively.

Finally we estimate the mass of the $\Psi$ particle. For this we expand
the potential in terms of the fields $\phi$ and $\psi$ and collect terms
which are second order in these fields.
We find
\begin{eqnarray}
V &=& \lambda (\Phi^*\Phi)^2 + \lambda_1(\Psi^*\Psi - \lambda_2^2
\Phi^*\Phi)^2\cr
&=& \lambda\left[\eta_1^2\phi_1^2+ \eta_2^2\phi_2^2 + 2\eta_1\eta_2\phi_1\phi_2+
 {1\over 2} (\eta_1^2+\eta_2^2) (\phi_1^2+\phi_2^2)\right]\cr
&+& \lambda_1\lambda_2^4 (\eta_1\phi_1 + \eta_2\phi_2)^2
-\delta\lambda_1\lambda_2^4 (\eta_1^2+\eta_2^2) (\phi_1^2+\phi_2^2)  \cr
&+& \lambda_1\big[ (\zeta_1\psi_1 + \zeta_2\psi_2)^2 + \delta\lambda_2^2
(\eta_1^2+\eta_2^2)  (\psi_1^2+\psi_2^2)\cr &-& 2\lambda_2^2
(\zeta_1\psi_1 + \zeta_2\psi_2)(\eta_1\phi_1 + \eta_2\phi_2) \big ]
+ \dots
\end{eqnarray}
where we have only displayed the quadratic terms in the fluctuations.
We now need to diagonalize the mass matrix. The form of these terms suggests
that we define the fields
\begin{eqnarray}
\phi_+ &=& \phi_1\cos\theta_0  + \phi_2\sin\theta_0 \cr
\phi_- &=& -\phi_1\sin\theta_0  + \phi_2\cos\theta_0\cr
\psi_+ &=& \psi_1\cos\theta_0  + \psi_2\sin\theta_0 \cr
\psi_- &=& -\psi_1\sin\theta_0  + \psi_2\cos\theta_0
\end{eqnarray}
where $\cos\theta_0 = \eta_1/\sqrt{\eta_1^2 + \eta_2^2}$ and
$\sin\theta_0 = \eta_2/\sqrt{\eta_1^2 + \eta_2^2}$.
In terms of the rotated fields the potential can be written as
\begin{eqnarray}
V &=& \Bigg\{\left[{3\over 2}\lambda + (1-\delta)\lambda_1\lambda_2^4\right]
\phi_+^2 + \left({\lambda\over 2} - \delta\lambda_1
\lambda_2^4\right) \phi_-^2\cr
&+& \lambda_1(1+3\delta)\lambda_2^2 
\psi_+^2 + \delta\lambda_1\lambda_2^2\psi_-^2 \cr
&-& 2\lambda_1\lambda_2^3(1+\delta)\phi_+\psi_+   \Bigg\}(\eta_1^2 + \eta_2^2).
\end{eqnarray}
We find that the states $\phi_+$ and $\psi_+$ mix with one another.
The mass matrix can be diagonalized and we find  the four states
\begin{eqnarray}
\psi'_+ &=& \psi_+\cos\theta_1 -  \phi_+\sin\theta_1 \cr
\phi'_+ &=& \psi_+\sin\theta_1  + \phi_+\cos\theta_1
\end{eqnarray}
$\psi_-$ and $\phi_-$
with mass squared eigenvalues, $2\lambda_1\lambda_2^2(\eta_1^2 + \eta_2^2) ,
3\lambda (\eta_1^2 + \eta_2^2), \lambda (\eta_1^2 + \eta_2^2)$
and $\lambda (\eta_1^2 + \eta_2^2) $ respectively. The mixing angle
$\theta_1<<1$ is approximately equal to $\lambda_2$.
We find one particle with relatively large mass of order
$2\lambda_1\lambda_2^2(\eta_1^2 + \eta_2^2)$. By adjusting $\lambda_2$ we
can choose this to be of order of 100 GeV and hence can model the Higgs
particle. The remaining three particles have very small masses. If we
gauge one of the U(1) symmetry then one of these particles will be eliminated
from the spectrum and will instead give rise to a massive gauge boson. The
theory then predicts two very light weakly coupled particles.

In this section we have considered a toy model which illustrates that we can
generate any mass scale by cosmological symmetry breaking in a theory with
pseudoscale invariance. The precise gauge group used $U(1)\times U(1)$ is not
essential for this purpose and the construction can be easily generalized to
the standard model. It is ofcourse important to check that quantum corrections
do not lead to acute fine tuning problems. We postpone this necessary check
to future research. We point out that an alternative to the construction in
this section is to simply gauge the pseudo-scale invariance
\cite{ChengPRL,Cheng,Hochberg,Wood,Wheeler,Feoli,Pawlowski,Nishino,Demir,Foot}. This eliminates the Higgs boson from the particle
spectrum and instead predicts a new vector boson with mass of the order of
Planck mass. 

\section{Conclusions}
We have shown that a slowly rolling solution to scalar field theories, leads
to breakdown of symmetries of the action. We call this phenomenon cosmological
symmetry breaking and show that it is intrinsically
different from spontaneous symmetry breaking. We argue that if we impose
pseudo-scale invariance on the action then it sets the cosmological
constant to zero both in the classical and the quantum theory.
The pseudo-scale invariance is also broken cosmologically, leading to
a slowly varying cosmological constant. We further show that cosmologically
broken pseudo-scale invariance can lead to a wide range of particle masses
and it appears possible to impose this symmetry on the full action of
fundamental particle physics.

{\bf Acknowledgements:} We thank S. D. Joglekar for useful discussions.

\end{document}